# Normalization, CWTS indicators, and the Leiden Rankings:

# Differences in citation behavior at the level of fields

Loet Leydesdorff [a] & Tobias Opthof [b,c]

**Abstract:** Van Raan *et al.* (2010; arXiv:1003.2113) have proposed a new indicator (*MNCS*) for field normalization. Since field normalization is also used in the Leiden Rankings of universities, we elaborate our critique of journal normalization in Opthof & Leydesdorff (2010; arXiv:1002.2769) in this rejoinder concerning field normalization. Fractional citation counting thoroughly solves the issue of normalization for differences in citation behavior among fields. This indicator can also be used to obtain a normalized impact factor.

[a] Amsterdam School of Communications Research (ASCoR), University of Amsterdam, Kloveniersburgwal 48, 1012 CX Amsterdam, The Netherlands; loet@leydesdorff.net; http://www.leydesdorff.net.
[b] Experimental Cardiology Group, Heart Failure Research Center, Academic Medical Center AMC, Meibergdreef 9, 1105 AZ Amsterdam, The Netherlands.
[c] Department of Medical Physiology, University Medical Center Utrecht, Utrecht, The Netherlands.

**Introduction**

In their rejoinder entitled "Rivals for the crown: Reply to Opthof & Leydesdorff," Van Raan *et al.* (2010) accepted our critique for the case of journal normalization (*CPP/JCSm*). However, a new indicator is proposed for field normalization (previously *CPP/FCSm*), called the "mean normalized citation score" (*MNCS*; cf. Lundberg, 2007).[4] In our opinion, this change does not sufficiently resolve the problems. Since the new indicator is proposed as another "crown indicator" (Waltman *et al.*, in preparation), it seems urgent to warn against and elaborate on the remaining problems. In addition to damaging evaluation processes at the level of individuals (PIs) and institutions, the "crown indicator" is also used by CWTS for the "Leiden Rankings," and flaws in it can therefore misguide policies at national levels.

Our previous critique focused on journal normalization because the journal indicator is analytically the simpler case. Journals provide clearly delimited units of analysis, while fields are compounded constructs. Formally, the CWTS indicators for journal and field normalizations could be considered as equivalent:

$$\frac{CPP}{JCSm} = \frac{((c_{1i}+c_{2i})+c_{3j}+c_{4k})}{(2J_i+J_j+J_k)} \quad (1)$$

$$\frac{CPP}{FCSm} = \frac{((c_{1i}+c_{2i})+c_{3j}+c_{4k})}{(2F_i+F_j+F_k)} \quad (2)$$

---

[4] The *MNCS* indicator is not to be confused with the existing indicator *NMCR* or Normalized Mean Citation Rate used by ECOOM in Leuven (Glänzel *et al.*, 2009, at p. 182). The *NMCR* of ECOOM-Leuven is equivalent to the old "crown indicator" *CPP/FCSm* of CWTS-Leiden.

In the first example (Eq. 1), the first two articles are published in the same journal (*i*) and normalized against the average of the citations of the reference set (in terms of, for example, document types and publication years) in this same journal, whereas the other two articles (*j* and *k*) are evaluated with reference to their respective reference sets. In Equation 2, the same is done for fields, but instead of the 8000+ journals of the ISI set, in this case the 221 ISI Subject Categories are used for the normalization. Note that various values for *F* will be equal in the case of different journals when the latter are subsumed under the same field or category. In other words, journal normalization is more finely grained than field normalization (Rafols & Leydesdorff, 2009), yet nevertheless the field-normalized one is considered by CWTS as their "crown indicator."

We have objected that the mean citation score can be properly normalized as follows respecting the arithmetic order of operations:

$$mean\ citation\ score = (\frac{c_{1i}}{J_i} + \frac{c_{2i}}{J_i} + \frac{c_3}{J_j} + \frac{c_4}{J_k})/4 \qquad (3)$$

In addition to the mean, the distribution (between the brackets in Eq. 3) also provides other statistics such as a standard deviation and the median.

Van Raan *et al.* (2010) counter-argue that the order of operations is just a convention which can be circumvented by placing brackets as in Equations 1 and 2. Our argument therefore is deemed "irrelevant." Of course, we understand that one can place brackets and thus force a change in the order of operations. However, changing the order of operations by using brackets also changes the meaning, traceability, and transparency of the indicators and evaluation outcomes (Spaan, 2010). In other words: 3/2 plus 2/3 is mathematically different from 5/5 and has another meaning as an indicator. By changing the order of operations, one loses the possibility to use statistics to determine whether observed differences are also significant.

Furthermore, the effects on the rankings at the level of individual researchers and research groups can be significant. We demonstrated this in the case of the research evaluation of the Academic Medical Center of the University of Amsterdam: it could be shown that one scholar who was rated by the Leiden evaluations as precisely at the world average was not significantly different in her[5] citation score from another scholar who belonged to the top group. The Leiden indicators fail to test for significance, and CWTS consequently has never provided error bars in the graphs.

As elsewhere (e.g., CWTS, 2008, at p. 7), Van Raan *et al*. (2010) provide references to Schubert & Glänzel (1983) or Glänzel (1992) to legitimate a difference of 0.2 as significant when unity is considered as "the world average." However, this value of 0.2 is not statistics, but a rule of thumb. In the interval between zero and one, 0.2 has a meaning different from above the world average because this interval is not limited to one to two.

Schubert & Glänzel (1983) based their reasoning on normal distributions (Glänzel, 2010). This reasoning can be used to estimate error in large sets (Glänzel, *personal communication*, 16 November 2009), but the estimator is insufficiently precise for evaluations of smaller sets. The alternative of bootstrapping mentioned by Van Raan *et al*. (2010) as another possible strategy makes the issues unnecessarily complex and has not yet been applied by CWTS. These references, in our opinion, disguise the fact that a statistics is missing from the CWTS evaluations hitherto, while our measures and the revised crown indicator, allow for the application of standard tests such as Kruskall-Wallis, as was demonstrated in our previous contributions (Opthof & Leydesdorff, 2010; Leydesdorff & Opthof, 2010).

---

[5] In order to respect anonymity, we use "her" as gender neutral.

Let us add that we were pleasantly surprised by the flexibility of CWTS to adapt its indicator to our criticism (Waltman *et al*., 2010). We note that some other centers (e.g., ECOOM in Leuven and ISSRU in Budapest) continue to use the quotient between the Mean Observed Citation Rates (MOCR) and the Mean Expected Citation Rates (MECR) as a main indicator (Glänzel *et al.,* 2009, at p. 182) using as an argument that the mean of the expectations is not a statistical function, but an expectation based on a set and therefore a real value (Glänzel, 2010; Glänzel, *personal communication*, March 18, 2010). It seems to us that this inference is only valid for large sets. In our opinion, institutional and *a fortiori* individual evaluations are to be tested using non-parametric statistics.

**Field normalization**

We focused on journal normalization because in the case of field normalization, one has two problems: the scientometric one of how to delineate fields and the statistical one of how to normalize. Journals are delineated units of analysis. Van Raan *et al*. (2010) have accepted our critique in the case of journal normalization,[6] but it seems to us that a new "crown indicator" is being hastily proposed for field normalization, with the mean normalized citation score (*MNCS*) taken as an alternative to *CPP/FCSm* (Equation 2). Like *CPP/FCSm*, *MNCS* is based on the ISI Subject Categories for weighing citation scores, as follows:

$$MNCS = (\frac{c_{1i}}{F_i} + \frac{c_{2i}}{F_i} + \frac{c_3}{F_j} + \frac{c_4}{F_k})/4 \qquad (4)$$

The weights of the different fields are derived from the average citation scores within each subject category. Equation 4 is analogous to our Equation 3, but now for fields instead of journals in the respective denominators. Although we now agree about the statistical normalization, this new "crown indicator" will inherit the scientometric problem of the previous one in treating subject categories as a standard for normalizing differences in citation behavior among fields of science.

1. The ISI Subject Categories were not designed for the scientometric evaluation, but for the purpose of information retrieval. Despite a strong denial by Van Raan *et al.* (2010) who formulate: "we are not aware of any convincing evidence of large-scale inaccuracies in the classification scheme of WoS," the subject categories lack an analytical base (Boyack *et al.*, 2005; Pudovkin & Garfield, 2002, at p. 1113n.; Rafols

---

[6] In his reaction to Opthof & Leydesdorff (2010), however, Moed (2010b) argues for using the old *CPP/JCSm* and *CPP/FCSm* ratios because at the level of aggregates (groups or oeuvres) distributions are less important, in his opinion. In our opinion, units of analysis can be aggregated and variables normalized; these two discussions are analytically different. However, distributions are crucial for testing the significance of observed differences, both at the level of individual cases and at the level of groups or their oeuvres. The "old crown indicators" could not provide us with distributions.

& Leydesdorff, 2009) and are not literary-warranted (Bensman & Leydesdorff, 2009; cf. Chan, 2005). As Opthof & Leydesdorff (2010) have explained in greater detail, alternative and far more precise classification schemes are available. Why not evaluate an academic hospital on the basis of the Medical Subject Headings (MeSH) of the bibliographic database MedLine, which are publicly available and compiled on a paper-by-paper basis (Bornmann *et al.*, 2009. at p. 98)?

2. If papers are published in journals which are attributed to several subject categories, CWTS chooses to weigh each category equally. This procedure generates artifacts in the rankings, since some journals are highly specialized (e.g., in cardiology) but nevertheless subsumed under a number of categories (for the purpose of information retrieval). The distinctions among categories are not based on multivariate analysis of the citation matrix among journals or weighted in terms of numbers of citations (Leydesdorff, 2006; Pudovkin & Garfield, 2002).

For example, the *Journal of Vascular Research* is attributed to the subject categories of "peripheral vascular disease" and "physiology," and the journal *Circulation* to "cardiac and cardiovascular systems," "hematology," and "peripheral vascular diseases"; whereas the *American Journal of Cardiology* is attributed only to "cardiac and cardiovascular systems." Scholars in these fields, however, publish and cite across such categorical divides.

3. The purpose of normalization at the field level is to control for differences in citation densities among fields. These differences are caused by differences in citation behavior among scholars in various fields of science. Mathematics, for example, is known to have a much lower citation density than the biomedical sciences (McAllister *et al*., 1983). However, the easiest way to capture this difference in citation behavior is by fractional counting in the citing articles at the article level. The level of aggregation for the benchmarking can then still be decided, for example, in terms of ISI Subject Categories.

For example, if an author in mathematics cites six references, each reference can be counted as 1/6 of overall citation, whereas a citation in a paper in biomedicine with 40 cited references can be counted as 1/40. This normalization thoroughly takes field differences into account and the results allow for statistical testing. Most importantly, this normalization is independent from a classification system and thus there is no indexer effect.

**Fractional citation counting as field normalization**

Moed (2010a) has proposed returning to fractional counting of citations in terms of the citing papers when recently constructing the so-called *SNIP* indicator for the *Scopus* database. (*SNIP* stands for "situated normalized impact per paper.") Small & Sweeny (1985) first applied fractional counting to co-citations in order to control for the noted differences of citation frequencies among fields of science.[7] Zitt & Small (2008) returned to fractional counting of the citations for solving the problem of the normalization, but use a formula (Equation 1 at p. 1858) in which averages are divided instead of averaging over a distribution of quotients. Moed (2010a) made the same mistake as previously when developing the Leiden indicators (Moed *et al*., 1995), namely, to first add up and then divide in both the numerator and denominator of the *SNIP*-indicator.

We elaborated this critique of the *SNIP* indicator elsewhere (Leydesdorff & Opthof, 2010), but we acknowledge that the original idea is fruitful because one can normalize on the basis of the citing articles directly for citation behavior, instead of using averages among rather arbitrarily delimited sets, such as fields of science operationalized as ISI Subject Categories or otherwise (e.g., Glänzel & Schubert, 2003; cf. Rafols & Leydesdorff, 2009). Let us thus turn our critique into a constructive proposal by showing the difference between the journal normalization contained in our previous contribution to this debate and the field normalization proposed here using the same seven PIs in our sample of the 232 scientists evaluated at the AMC.

---

[7] Fractional attribution of coauthored publications to authors was first proposed by Price & Beaver (1966). The idea of normalizing by fractionating the citation impact proportionately was developed by Narin (1976) and Pinski & Narin (1976), but elaborated by these authors in the different direction of so-called influence weights.

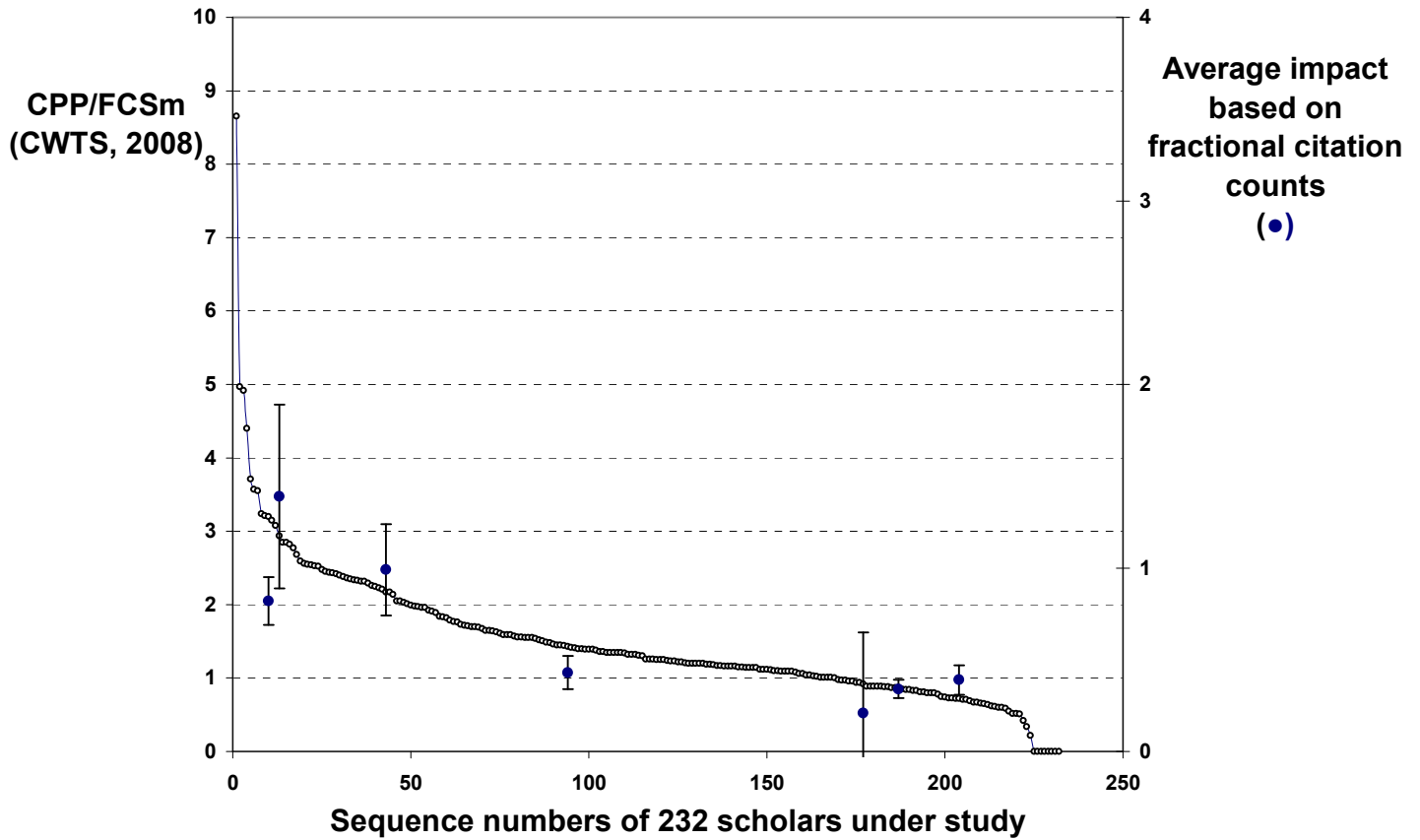

**Figure 1**: Ranking of 232 AMC scientists in terms of their *CPP/FCSm* (that is, field-normalized) according to the CWTS (2008); ● average citation impact using fractional citation counting for seven of these scientists.

Figure 1 can be compared with Figure 1 in Opthof & Leydesdorff (2010), but the present ranking is based on the Leiden field normalization (*CPP/FCSm*) instead of the journal normalization (*CPP/JCSm*). (As noted, we did not obtain values for the new indicator *MNCS*, but instead use the *CPP/FCSm* provided by CWTS (2008) for the comparison.) One can observe by visual inspection of the two graphs that the differences in the normalized citation scores based on fractional counting are larger in this case than the correction in the previous case of journal normalization despite differences in the scales.[8]
For us, this result does not come as a surprise because of the problem of the disturbing field delineations. In our opinion, the field-normalized "crown indicators" (both *CPP/FCSm* and *MNCS*) are for this reason less reliable than the journal-normalized indicator of CWTS (*JCSm*).

| | *Bibliometric data* | | | *Journal normalized* | | *Field normalized* | | |
|---|---|---|---|---|---|---|---|---|
| Rank | $\Sigma p_i$ | $\Sigma c_i$ | $Avg(c/p)$ | Mean citation score (previous study) | CPP/ JCSm (CWTS, 2008) | $\Sigma c_f$ (this study) | $Avg(c_f)$ (this study) | CPP/FCSm (CWTS, 2008) |
| 6 | 23 | 891 | 38.74 (± 13.67) | 2.03 (± 0.55) | 2.18 | 31.95 | 1.39 (± 0.50) | 2.94 |
| 14 | 37 | 962 | 26.00 (± 4.09) | 1.74 (± 0.19) | 1.86 | 30.32 | 0.82 (± 0.13) | 3.20 |
| 26 | 22 | 567 | 25.77 (± 5.78) | 1.54 (± 0.23) | 1.56 | 21.74 | 0.99 (± 0.25) | 2.17 |
| 117 | 32 | 197 | 6.16 (± 1.30) | 1.50 (± 0.29) | 1.00 | 6.83 | 0.21 (± 0.44) | 0.92 |
| 118 | 37 | 402 | 10.86 (± 2.21) | 0.93 (± 0.13) | 1.00 | 16.08 | 0.43 (± 0.09) | 1.43 |
| 206 | 65 | 647 | 9.96 (± 1.57) | 0.91 (± 0.11) | 0.58 | 21.90 | 0.34 (± 0.05) | 0.87 |
| 223 | 32 | 354 | 11.06 (± 1.74) | 0.78 (± 0.12) | 0.43 | 12.40 | 0.39 (± 0.08) | 0.72 |
| | | | | Spearman $\rho > 0.99$; $p < 0.01$ | | | Spearman $\rho = 0.75$; *n.s.* | |

**Table 1**: The effects of different normalizations on values and ranks

Table 1 quantifies these effects. The journal normalizations in the middle of this table correspond to the figures provided in Table 4 of Opthof & Leydesdorff (2010). Whereas the journal normalizations correlate highly in terms of their rank ordering (Spearman's $\rho > 0.99$; $p < 0.01$) despite considerable differences at the level of individual scores, the two field normalizations—this study *versus* CWTS (2008)—no longer correlate even when using $p < 0.05$. One can expect values for the *MNCS* to be highly correlated with those for *CPP/FCSm* (Van Raan, 2010, at slides 31-34; Van Raan *et al.*, 2010, at p. 5; Waltman *et al.*, 2010), and therefore not with the weighted citation impact based on fractional counting.

---

[8] Different from Figure 1 in Opthof & Leydesdorff (2010), the scales are now unequal because the average impact is based on fractional counting and *CPP/FCSm* on whole-number counting. A world average would require normalization against a fractionally counted impact factor for each journal. One would thus be able to combine field and journal normalization and develop what could perhaps be considered as "crown indicator."

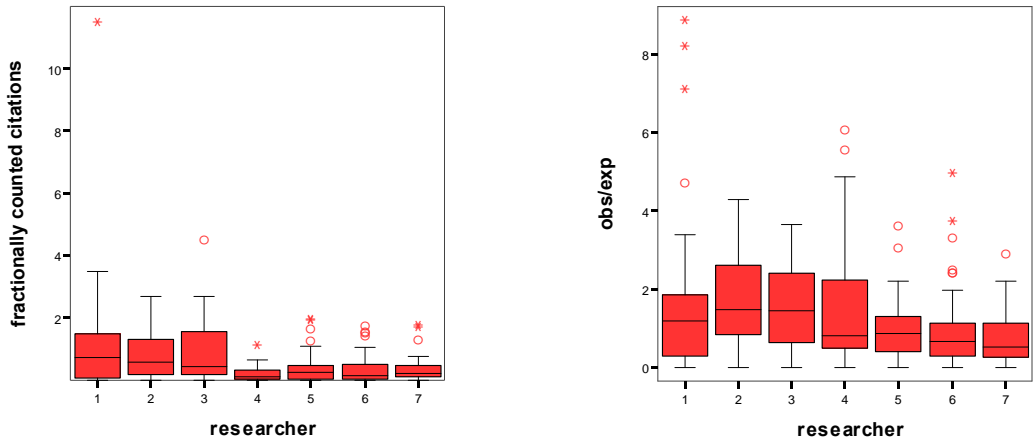

**Figure 2**: Boxplot of fractionally counted citations (left; this study) versus normalized citation rates (right; Opthof & Leydesdorff, 2010) for seven researchers in the AMC. (NB. Instead of means, medians are indicated as the lines in the box plots; the means are provided in Table 1.)

Figure 2 shows box plots of the distributions of fractionated citations (left) compared to our previous results (right) based on observed versus expected citation rates (Opthof & Leydesdorff, 2010). Whereas we found, for example, no significant differences between the first and fourth authors in the case of journal normalization using a *post-hoc* test with Bonferroni correction, the first, second, and third author are now a homogenous set. Furthermore, the third author's fractioned citation profile is significantly different from the fourth (using the Tukey test).[9] Using fractional counting for the normalization at the field level, one would hence be warranted to distinguish two groups among these seven researchers.

---

[9] The Bonferroni correction is often considered as too conservative. The equivalent Tukey test in SPSS is corrected for multiple comparisons (in addition to dyadic ones).

Note that by using fractional citation counts one abandons the notion of a world average as a standard for a field of science. Given the overlaps among fields, such a general standard is, in our opinion, sociologically unwarranted. By using fractional citation counts, however, one can benchmark against any reference set including the ones subsumed under the 221 ISI Subject Categories or the 60 subfields distinguished by ECOOM (Glänzel & Schubert, 2003; Glänzel *et al*. 2009). An advantage is that one can then use standard statistics to determine whether the performance above or below this "world average" is also significant. A further extension to non-parametric statistics as advocated by Bornmann (2010; cf. Plomp, 1990; Leydesdorff, 1990 and 1995) remains possible.

**Conclusion and discussion**

In the previous paper, we provided the corrected normalization for journals as intended by the *CPP/JCSm*, and in this paper we have extended our analysis with an explanation of how to normalize at the level of fields of science in terms of differences in citation behavior using fractional citation counts. In our opinion, this normalization accords with the intention behind the "crown indicators" of CWTS, but the latter assume the ISI Subject Categories to cover these differences in citation behaviors. We showed that the two normalizations lead to even significantly different results in our sample of seven researchers (Figure 2). These differences between field-normalization and journal-normalization accord with notions of science as global elite structures (Merton, 1968 and 1973; Whitley, 1984). Journals can be expected to organize more specific hierarchies; for example, by gate-keeping (e.g., Bollen *et al.*, 2005; Doreian & Farraro, 1985; Zsindely *et al.*, 1982).

Our measure of fractional counting can be generalized as normalization for any differences in citation behavior among citing authors (Small & Sweeney, 1985). Note that authors can also differ in terms of their publication behavior, and that these differences can be systematic among fields of science. However, differences in publication behavior cannot be captured by a citation indicator (Ulf Sandström, *personal communication*, March 5, 2010).

Fractional citation counting is simple and elegant. The resulting distributions can be analyzed statistically; error bars consequently can be indicated in the graphical results. The importation of indexer-based and potentially biased schemes of classification is no longer necessary. In another context (Leydesdorff & Opthof, 2010), we show that this measure can also be used to normalize the impact of journals by considering the citable issues in the denominator of the ISI-Impact Factor as a document set (in the years $t - 1$ and $t - 2$) which can be counted fractionally in terms of citations in the year $t$ (in the numerator). Thus, the measure is very general. As noted, we consider the Bonferroni correction *ex post* and its further refinements (e.g., the Tukey and Scheffé tests) as appropriate for testing significance among different sets.[10] These tests are available in statistical packages such as SPSS.

---

[10] Additionally, a robust test of the equality of the means is provided by the Welch statistics (Glänzel, *personal communication*, March 18, 2010). However, the means among these seven authors were significantly equal. In our opinion, one can also use Kruskall-Wallis for this purpose.

Van Raan, A. F. J. (2010). Quantitative Studies Studies of Science: Outlook. Paper presented at the *3rd Intl. Conference of the European Network for Indicators Designers ENID*, Paris, 5 March 2010.

Van Raan, A. F. J., Van Leeuwen, T. N., Visser, M. S., Van Eck, N. J., & Waltman, L. (2010). Rivals for the crown: Reply to Opthof and Leydesdorff. *Journal of Informetrics,* 4(3), in print.

Waltman, L., Van Eck, N. J., Van Leeuwen, T. N., Visser, M. S., & Van Raan, A. F. J. (2010). Towards a New Crown Indicator: Some Theoretical Considerations. *In preparation*.

Whitley, R. D. (1984). *The Intellectual and Social Organization of the Sciences*. Oxford: Oxford University Press.

Zitt, M., & Small, H. (2008). Modifying the journal impact factor by fractional citation weighting: The audience factor. *Journal of the American Society for Information Science and Technology, 59*(11), 1856-1860.

Zsindely, S., Schubert, A., & Braun, T. (1982). Editorial Gatekeeping Patterns in International Science Journals—A New Science Indicator. *Scientometrics,* 4(1), 57-68.17